\documentclass[aps,prb,twocolumn,superscriptaddress,unsortedaddress,amsmath,amssymb,showpacs,subeqn]{revtex4}
\usepackage{graphicx}
\usepackage{epsfig}
\usepackage{natbib}
\usepackage{color}

\begin{document}

\title{\parbox{18cm}{Distinguishing coherent from incoherent charge transport in linear triple quantum dots}}

\author{L.~D.~Contreras-Pulido}
\affiliation{Instituto de F\'{i}sica, Universidad Nacional Aut\'{o}noma de M\'{e}xico, Apartado Postal 20-364, 01000 Mexico City, Mexico}

\author{M.~Bruderer}
\affiliation{Institute of Theoretical Physics, Albert-Einstein Allee 11, Ulm University, 89069 Ulm, Germany}

\date{\today}

\begin{abstract}
A fundamental question in quantum transport is how quantum coherence influences charge transfer through nanostructures. We address this issue for linear triple quantum dots by comparison of a Lindblad density matrix description with a Pauli rate equation approach and analyze the corresponding zero-frequency counting statistics for coherent and sequential charge tunneling, respectively. The impact of decaying coherences of the density matrix due to dephasing is also studied. Our findings reveal that the sensitivity to coherence shown by shot noise and skewness, in particular in the limit of large coupling to the drain reservoir, can be used to unambiguously evidence coherent processes involved in charge transport across triple quantum dots. Our analytical results are obtained by using the characteristic polynomial approach to full counting statistics.
\end{abstract}


\maketitle

\section{Introduction}

Coherent superpositions of states are one of the fundamental aspects of quantum mechanics. They are an important resource for quantum information processing, quantum transport and quantum metrology.
A tunable platform for the manipulation of coherent quantum states is provided by arrays of semiconductor quantum dots (QDs).
Double quantum dot (DQD) systems have allowed the observation of superpositions of electronic states via coherent charge oscillations\cite{hayashi03,shinkai09} and sharp resonances in the current across the system.\cite{vanderwiel03} DQDs have been analyzed to a great extent, unveiling intriguing transport phenomena such as Pauli spin blockade\cite{ono02} or the Kondo effect.\cite{kastner97} They are also promising candidates for quantum information processing due to the coherent manipulation of charge or spin degrees of freedom.\cite{loss98,vanderwiel03,hanson07}

The coupling of three QDs represents the next level of complexity toward the development of nanoelectromechanical systems,\cite{flindt04,villavicencio08} and quantum information and simulation architectures.\cite{taylor10,granger12b,taylor13,delbecq16,vandersypen13c} The exceptional control and tunability recently achieved in triple quantum dots (TQDs)\cite{gaudreau06,gaudreau07,granger10,tarucha13} allowed the experimental verification of long-distance tunneling (LDT) of charge and spin between peripheral QDs,\cite{busl13,vandersypen13,sanchez14} opening the avenue to investigate different coherent phenomena. LDT and quantum interferences may determine the transport properties in the linear TQD, as demonstrated by the superexchange blockade\cite{sanchez14b} and by photon-assisted transitions in ac-driven systems.\cite{vandersypen15,sanchez15,sanchez16}

Motivated by the recent experimental progress, we analyze in this paper the zero-frequency counting statistics of charge transport through a serially coupled TQD attached to electronic reservoirs, as schematically depicted in Fig.~\ref{fig_TQD}. Our main goal is to find the parameter conditions for which fluctuations in charge transport make it possible to distinguish between coherent and incoherent transport. For this purpose, we compare a coherent with a fully incoherent description of the system. In the former, we assume that the QDs are tunnel-coupled and use a density matrix (DM) approach to characterize the system, including both diagonal (occupations) and non-diagonal (coherences) elements. The incoherent model corresponds to a rate equation description without coherences. In addition, as an intermediate model, we consider the effect of pure dephasing, induced by a phonon environment, on the tunnel-coupled TQD.\cite{contreras14}

\begin{figure}[t]\label{fig_TQD}
        \includegraphics[width=0.99\linewidth,clip]{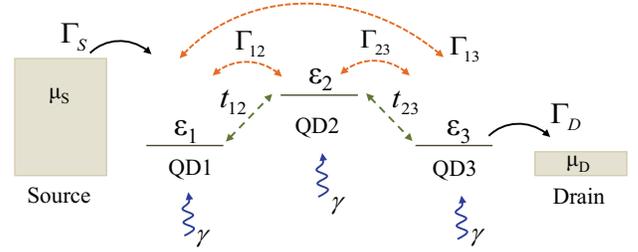}
\caption{(Color online) TQD in series connected with rates $\Gamma_S$ and $\Gamma_D$ to source and drain reservoirs at chemical potentials
$\mu_{S}$ and $\mu_{D}$. The coherent description of the system considers tunnel-coupled adjacent QDs with amplitudes $t_{12}$ and $t_{23}$. Incoherent transport is described by electronic transitions between the QDs with rates $\Gamma_{ij}$, including a direct transition between QD1 and QD3. An intermediate model with decaying coherences includes pure dephasing acting on the three QDs with rate $\gamma$.}
\label{fig_TQD}
\end{figure}

In order to analyze the effects of coherence we focus on the steady state transport properties of the TQD: the average current, its fluctuations in form of the Fano factor and the skewness as the first correlator beyond shot noise. The fully coherent and fully incoherent descriptions yield the same average current, however, differences in the statistics of higher-order current correlations allow to establish the coherent nature of charge transport: In the limit of large coupling to the drain $\Gamma_D$, the Fano factor for the coherent model can be super-Poissonian. In contrast, the fully incoherent model and the description including pure dephasing always exhibit sub-Poissonian noise. This difference in the current fluctuations stems from the fact that only the fully coherent model undergoes an equivalent in transport of the quantum Zeno effect. In accord with this interpretation, the skewness also allows to distinguish between coherent and incoherent descriptions for sufficiently large $\Gamma_D$ and finite detuning.

Different protocols are oriented toward verifying quantum coherence in quantum systems, ranging from Leggett-Garg inequalities to quantum state tomography.\cite{white01,nori12,huelga15} In the context of quantum transport, non-linear spectroscopy has been applied to assess coherence on exciton transfer in coupled systems\cite{richter15,mermillod16} while Leggett-Garg-type inequalities\cite{lambert10,nori12,emary14} and Landau-Zener-St\"{u}ckelberg interference\cite{vandersypen13} have been used to test the coherent behavior of electronic transport in nanostructures. Theoretical studies in DQDs have demonstrated the sensitivity to quantum coherence of the current and high-order cumulants.\cite{kiesslich06,kiesslich07,mukamel08,you13} In the same vein, we use here the counting statistics in order to verify the ``quantumness'' of charge transport in TQDs.

Our proposal relies on the experimental control of real-time detection of single-electron tunneling achieved in a linear TQD.\cite{vandersypen13} Real-time charge detection techniques\cite{rimberg03,delsing05,gustavsson06} paved the way for experimental measurement of the full counting statistics\cite{levitov93,levitov96,bagrets03} (FCS) of charge transport in diverse QD systems under out-of-equilibrium conditions.\cite{fujisawa04,fujisawa06,haug07,gustavsson06,gustavsson09,belzig05,emary07,ensslin07,schaller09,flindt11,ramon04,emary09c,belzig15,braggio06,flindt10,marcos11} Striking experiments addressed high-order cumulants\cite{flindt09,haug10} and finite-frequency statistics\cite{flindt12} in a single QD. The counting statistics for DQDs has been accessed for the second-order fluctuations or shot noise,\cite{haug06,kiesslich07} and theoretical studies of high-order cumulants reveal information about quantum coherence\cite{kiesslich06,kiesslich07,liu14} and detector-induced backaction.\cite{gaspard13,you13} For TQDs, shot noise has been analyzed mainly for triangular configurations,\cite{kuzmenko06,michaelis06,emary07b,sun07,emary09,dominguez11} while for the linear array it was studied in presence of dephasing.\cite{contreras14}

The paper is structured as follows: In Section~\ref{model} we introduce the theoretical models for the linear TQD in a transport configuration. We first present the coherent model in terms of a Lindblad master equation, describing the dynamics of the reduced density matrix of the TQD. From the quantum master equation we derive a Pauli rate equation for the populations of the electronic sites, defining our fully incoherent model. As an intermediate description of the system, we consider the effects of decaying coherences of the density matrix by introducing pure dephasing. In Section~\ref{transport}, we present our results for the counting statistics of the charge transport for each model. We center our attention on the behavior of the Fano factor and skewness with and without coherences. In Section~\ref{exp}, we suggest a procedure for the experimental verification of the presence of quantum coherence in the TQD. This section is followed by the Conclusions.

\section{Different models and method}\label{model}

The TQD consists of three single-level quantum dots arranged in series and connected to source and drain electron reservoirs (cf.~Fig.~\ref{fig_TQD}). Similar to Ref.~\onlinecite{contreras14}, we assume strong Coulomb blockade such that the TQD can be occupied with at most one extra spinless electron. The relevant states of the system are thus the empty state $|0\rangle$ and the single-particle states $\{|1\rangle$, $|2\rangle$, $|3\rangle\}$, where $|i\rangle$ describe the $i$-th QD being occupied. We consider the large bias voltage regime, with all the electronic states inside the conduction window, and tunneling of electrons is allowed from the source to QD1 and from QD3 into the drain, i.e., transport is unidirectional.

\subsection{Fully coherent model}

For the fully coherent description of the TQD it is considered that neighboring dots are tunnel coupled, depicted by green dashed arrows in Fig.~\ref{fig_TQD}. The Hamiltonian representing the array is then (we take $\hbar=1$ throughout the paper)
\begin{equation}
	H = (t_{12}d_1d_2^{\dagger}+t_{23}d_2d_3^{\dagger}+ {\rm h.c.}) + \sum_{i=1}^3\varepsilon_{i}d_{i}^{\dagger}d_{i}\,,
\end{equation}
where $t_{ij}$ are tunneling couplings between QDs and $d_{i}^{\dagger}$ ($d_{i}$) is the creation (annihilation) operator for an electron in the $i$-th QD.

In the large bias voltage regime and for strong Coulomb blockade, the time evolution of the reduced DM $\rho(t)$ of the TQD in a transport configuration reads\cite{contreras14}
\begin{equation}\label{rhot}
\begin{split}
\dot{\rho}(t)&=\mathcal{L}_0\rho(t)\\
&=-i[H,\rho(t)]+\Gamma_S\mathcal{D}(d_1^{\dagger})\rho(t)+\Gamma_D\mathcal{D}(d_3)\rho(t)\,,
\end{split}
\end{equation}
with the dissipators of Lindblad form $\mathcal{D}(A)\rho=A\rho A^{\dagger}-\frac{1}{2}A^{\dagger}A\rho-\frac{1}{2}\rho A^{\dagger}A$.\cite{breuer,rivas}
The superoperators $\mathcal{D}(d_1^{\dagger})$ and $\mathcal{D}(d_3)$ describe irreversible tunneling of electrons from the source and into the drain, respectively, with rates $\Gamma_S$ and $\Gamma_D$. In the wide-band approximation and infinite bias limit the rates are energy independent and, moreover, the Born-Markov approximation with respect to the coupling to the reservoirs is essentially exact.\cite{brandesrep,timm08} In the chosen basis, the reduced DM $\rho$ in Eq.~(\ref{rhot}) contains diagonal (occupation probabilities for each QD) and non-diagonal (coherences) elements. Detailed expressions for the elements $\dot{\rho}_{ij}(t)$ are explicitly given in Appendix~\ref{appss}.

\subsection{Incoherent model}

For incoherent transport we study the evolution of the linear TQD with only diagonal elements of the density matrix. In a similar spirit, transport in DQDs without coherences has been analyzed in Refs.~\onlinecite{kiesslich06} and \onlinecite{kiesslich07}, however we focus here on the parameters for which it is possible to discern the nature of transport.
From the fully coherent DM, Eq.~(\ref{rhot}), we derive an analogous classical model for TQDs in form of a Pauli rate equation (RE) for the occupation probabilities of the electronic states, arranged in the vector $\mathbf{P}=(\rho_{11}, \rho_{22}, \rho_{33}, \rho_{44})^{T}$. The RE (derived in Appendix~\ref{appss}) has the form
\begin{equation}\label{Pt}
\dot{\mathbf{P}}(t)=\mathcal{L}_I\mathbf{P}(t),
\end{equation}
with the generator of the dynamics $\mathcal{L}_I$ given by
\begin{equation}\label{Lich}
\mathcal{L}_I = \left(
\begin{array}{cccc}
- \sigma_1 & \Gamma_{12} & \Gamma_{13} & \Gamma_{S}\\
\Gamma_{21} & - \sigma_2 & \Gamma_{23} & 0\\
\Gamma_{31} & \Gamma_{32}& - \sigma_3 - \Gamma_{D} & 0\\
0 & 0 & \Gamma_{D} & -\Gamma_{S},
\end{array}
\right)
\end{equation}
and $\sigma_j = \sum_{i\neq j}\Gamma_{ij}$.
The rates $\Gamma_{ij}=\Gamma_{ji}$ are symmetric and describe incoherent transitions between electronic states, depicted by dotted lines in Fig.~\ref{fig_TQD}. Their dependence on the system parameters is given by
\begin{eqnarray}\label{rates_ich}\nonumber
\Gamma_{12} &=& 4 t_{12}^2 t_{23}^2 \Gamma _D \left[\Gamma _D^2+4 t_{12}^2-4 t_{23}^2+4 \left(\varepsilon _{23}^2+x\right)\right]/\Delta\,,\\[3pt]
\Gamma_{13} &=& 16 t_{12}^2 t_{23}^2 \Gamma _D \left(t_{23}^2-x\right)/\Delta\,,\\[3pt]\nonumber
\Gamma_{23} &=& 4 t_{23}^2 \Gamma _D \left[\varepsilon _{12}^2 \Gamma _D^2+4 t_{12}^2 \left(\varepsilon _{12}^2-t_{23}^2+x\right)+4 y^2\right]\Delta\,,
\end{eqnarray}
with $\varepsilon_{ij}=|\varepsilon_{i}-\varepsilon_{j}|$ the difference between energy levels, $x=\varepsilon _{12} \left(\varepsilon _{13}+\varepsilon _{23}\right)$, $y=t_{23}^2-\varepsilon _{12} \varepsilon _{13}$ and
\begin{equation}\nonumber
\begin{split}
\Delta &= \varepsilon _{12}^2 \Gamma _D^4+4 \Gamma _D^2 \left[\varepsilon _{12}^2 \left(\varepsilon _{23}^2+2 t_{12}^2\right)+y^2\right]\\
&+16 \left(\varepsilon _{12} t_{12}^2+\varepsilon _{23} y\right)^2\,.
\end{split}
\end{equation}
Note that the rates $\Gamma_{ij}$ are independent of the coupling to the source $\Gamma_S$ and include direct transitions between the peripheral dots QD1 and QD3. Expressions~(\ref{rates_ich}) result from eliminating the coherences from the DM and are therefore not equivalent to rates obtained from Fermi's golden rule.

We mention that there is not a unique incoherent description of the TQD, i.e., different generators $\mathcal{L}_I$ (or equivalently rates $\Gamma_{ij}$) in the RE may yield identical results in the steady-state.\cite{bruderer14} Nevertheless, in the remainder of the paper we exclusively use the RE~(\ref{Pt}) with the rates in (\ref{rates_ich}), referred to as \textit{the} incoherent model.

\subsection{Pure dephasing}

We model charge transport including the effect of decaying coherences of the DM by introducing pure dephasing on the electronic sites of the TQD. We consider electron-phonon interactions as the main source of dephasing,\cite{fedichkin04,hu05,fujisawa06b} and neglect relaxation processes by significantly detuning the central QD. (For details about the physical origin of pure dephasing in lateral QDs, cf.~Ref.~\onlinecite{contreras14} and references therein).

Pure dephasing affects the DM in Eq.~(\ref{rhot}) through an additional Lindblad term such that the time evolution of the DM in presence of dephasing reads
\begin{equation}\label{rhodt}
\dot{\rho}(t) = \mathcal{L}_\phi\rho(t) = \mathcal{L}_0\rho(t) + \gamma\sum_i\mathcal{D}(n_i)\rho(t)\,,
\end{equation}
with $n_i = d^{\dagger}_{i}d_{i}$. Here, we have assumed that the dephasing rate $\gamma$, which causes and exponential decay of the coherences of $\rho(t)$, is equal for all three QDs.

\subsection{FCS of charge transport}

The effect of coherence on the steady-state transport properties of the linear TQD is studied by analyzing the average current, the zero-frequency noise and the skewness. The central quantity in zero-frequency FCS is the probability distribution $p(N)$ for the number of charges $N$ transferred into the drain reservoir within a long time interval $\Delta\tau$. The distribution $p(N)$ is conveniently characterized by its cumulants $C_k$.

We use the characteristic polynomial approach\cite{bruderer14} to determine the time-scaled cumulants $c_k = C_k/\Delta\tau$, and from them we obtain the properties of interest: The average current $I=ec_1$ and the shot noise $S = 2ec_2$ (with $e$ the electron charge). The noise strength is characterized by the Fano factor $F^{(2)}=S/2eI$, which determines if the distribution is super-Poissonian, $F^{(2)}>1$, or sub-Poissonian, $F^{(2)}<1$. The third cumulant is related to the skewness of the distribution $p(N)$, parametrized here by the Fano factor $F^{(3)}=c_3/c_1$.
Analytical expressions for the FCS obtained for the aforementioned models are explicitly given in Appendix~\ref{appFCS}.

\begin{figure}[t]
\includegraphics[width=1\linewidth,clip]{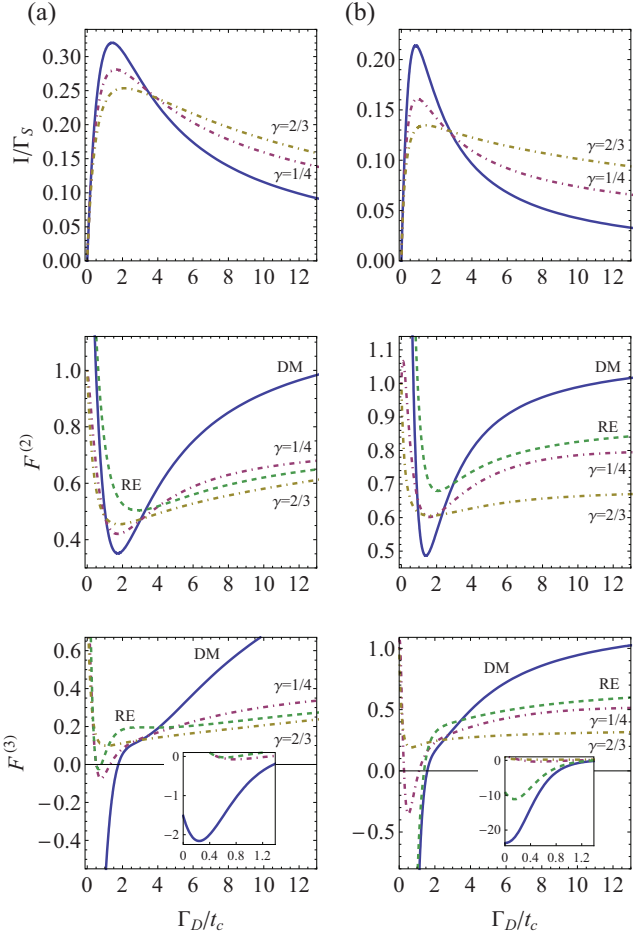}
\caption{(Color online) Coherent and incoherent transport through the TQD in the steady state. Average current $I$, Fano factor $F^{(2)}$ and normalized skewness $F^{(3)}$ as a function of $\Gamma_D$, for the fully coherent DM (solid line), incoherent RE (dashed), and DM with dephasing (dash-dotted lines). The detuning $\varepsilon$ of the central QD is set to (a) $\varepsilon = 2 t_c$ and (b) $\varepsilon= 4 t_c$. The insets in the bottom panel show details of the skewness in the regime $\Gamma_D\ll t_c$. (We use $\gamma=1/4 t_c$, $\gamma=2/3 t_c$ and $\Gamma_S=1/2 t_c$).}
\label{fig_cumulants}
\end{figure}

\section{Comparison of coherent and incoherent transport}\label{transport}

For clarity, in the reminder of the paper we consider identical tunnel couplings between neighboring QDs, $t_{ij}=t_c$. The TQD is assumed to be in a $\Lambda$- or $V$-type configuration with QD1 and QD3 in resonance, $\varepsilon_1=\varepsilon_3$. The energy difference between the central and outer dots is then determined by the detuning $\varepsilon=|\varepsilon_1-\varepsilon_2|$.

\subsection{Fully coherent versus incoherent model}\label{coherent}

The steady-state occupation probabilities obtained from the fully coherent DM in Eq.~(\ref{rhot}) and the incoherent RE~(\ref{Pt}) are equal and therefore both models yield the same stationary average current
\begin{equation}\label{currcoh}
I_0 = \frac{4 t_c^4\Gamma_S\Gamma_D}{4t_c^4 (3\Gamma_S+\Gamma_D)+2t_c^2\Gamma_S\Gamma_D^2+\Gamma_S\Gamma_D^2\varepsilon^2}\,.
\end{equation}
This feature stresses the importance of understanding the role of the coherences in electronic transport by analyzing the current fluctuations and higher-order cumulants.
The current~(\ref{currcoh}) is shown in Fig.~\ref{fig_cumulants} as a function of the incoherent coupling to the drain $\Gamma_D$ and for different energy detunings $\varepsilon$ (parameters are in units of $t_c$). We observe that $I_0$ is peaked in the regime of small to intermediate coupling to the drain, $\Gamma_D\lesssim 2 t_c$, and is suppressed as $\Gamma_D$ increases.

For the fully coherent model~(\ref{rhot}), we can interpret these results in terms of the dynamics of the charge:\cite{contreras14} For large detuning of the central QD and $\Gamma_D\lesssim 2t_c$, coherent LDT of charge from QD1 to QD3 provides the main mechanism for transport. On the other hand, for large coupling to the drain $\Gamma_D\gg t_c$, the system undergoes a counterpart of the quantum Zeno effect in quantum transport.\cite{gurvitz98,chen04} Large values of $\Gamma_D$ are considered to be equivalent to a continuous measurement of the occupation of QD3, projecting the charge into the subsystem QD1 and QD2, and therefore blocking the current.
For both limiting cases, the dynamics of the TQD is reduced to an effective DQD, formed by QD1 and QD3 for $\Gamma_D\lesssim 2t_c$ and by QD1 and QD2 for $\Gamma_D\gg t_c$.\cite{contreras14}

Both LDT and the Zeno regimes rely on the coherence of the system and are reflected in the behavior of the off-diagonal elements $\rho_{ij}^s$ of the steady-state DM, provided in Appendix~\ref{appcoh}. The coherence $\rho_{13}^s$, indicating LDT between the peripheral dots of the TQD, is the dominating contribution for the total coherence of the system in the regime $\Gamma_D\lesssim 2 t_c$. For $\Gamma_D\gg t_c$, the element $\rho_{12}^s$ accounts for most of the coherence, as a signature of the trapped coherent evolution of the electron induced by the Zeno effect.

Let us compare these findings with the mechanisms for fully incoherent transport. In the regime $\Gamma_{D}\lesssim 2 t_c$, the transition rate $\Gamma_{13}$ between QD1 and QD3 accounts for most of the charge transfer. Thus, similar to the coherent case, direct transitions between the outermost QDs dominate transport. In contrast, for intermediate values of $\Gamma_D$  and for $\Gamma_D\gg t_c$, the rates $\Gamma_{12}$ and $\Gamma_{23}$ determine most of the electronic transitions, indicating sequential transport. The incoherent transition rates $\Gamma_{ij}$ in Eqs.~(\ref{rates_ich}) are further discussed in Appendix~\ref{appich}.

We can therefore conclude that the underlying mechanism yielding a current resonance in the regime $\Gamma_D\lesssim 2 t_c$ is similar for coherent~(\ref{rhot}) and incoherent~(\ref{Pt}) descriptions of the TQD and is related to charge transfer from QD1 to QD3. However, for sufficiently large coupling to the drain $\Gamma_D\gg t_c$, the physical mechanisms involved in transport are remarkably different.

Unlike the current, the high-order cumulants are sensitive to quantum coherence and therefore different for the two models. The coherent Fano factor $F^{(2)}_0$ and the incoherent one $F^{(2)}_I$ read (cf.~Appendix~\ref{appFCS} for details).
\begin{widetext}
\begin{equation}\label{Fcoh}
F^{(2)}_{\textrm{0}} = 1+\frac{2 t_c^2 \Gamma _S \left(4 t_c^4 \left[4 t_c^2 \left(\Gamma _S-3 \Gamma _D\right)-\Gamma _D^2 \left(7 \Gamma _S+2 \Gamma _D\right)\right]+\left[16 t_c^4 \Gamma _S+\Gamma _S\Gamma _D^4 -4 t_c^2 \Gamma _D^2 \left(5 \Gamma _S+\Gamma _D\right)\right]\varepsilon ^2 \right)}{\left[4 t_c^4 \left(3 \Gamma _S+\Gamma _D\right)+2 t_c^2  \Gamma _S \Gamma _D^2+\Gamma _S\Gamma _D^2 \varepsilon ^2 \right]^2}\,,
\end{equation}
\begin{equation}\label{Fich}
F^{(2)}_{\textrm{I}} = 1+\frac{2 t_c^2 \Gamma _S \left(4 t_c^4 \left[4 t_c^2 \left(\Gamma _S-3 \Gamma _D\right)-\Gamma _D^2 \left(\Gamma _S+2 \Gamma _D\right)\right]+\left[16 t_c^4 \Gamma _S- \Gamma _S\Gamma _D^4 -4 t_c^2 \Gamma _D^2 \left(\Gamma _D+\Gamma _S\right)\right]\varepsilon ^2\right)}{\left[4 t_c^4 \left(3 \Gamma _S+\Gamma _D\right)+2 t_c^2 \Gamma _S \Gamma _D^2+\Gamma _S \Gamma _D^2 \varepsilon ^2\right]^2}\,.
\end{equation}
\end{widetext}
Both Fano factors can be sub- or super-Poissonian depending on the relative values of the energy detuning $\varepsilon$, the interdot tunneling coupling $t_c$ and the couplings to the reservoirs $\Gamma_S$ and $\Gamma_D$.

Due to the intricate dependence of Eqs.~(\ref{Fcoh}) and~(\ref{Fich}) on the parameters of the TQD, we explore first the ratio between the Fano factors $F^{(2)}_{\textrm{I}}/F^{(2)}_0$, such that significant deviations from unity indicate differences between coherent and incoherent transport.
This ratio is mapped out in Fig.~\ref{fig_mapF} as a function of $\Gamma_D$ and $\varepsilon$ for a representative value of $\Gamma_S$. We observe two complementary regions in which the relative current fluctuations are enhanced and reduced owing to coherence. The former corresponds to the regime $\Gamma_D\lesssim 2 t_c$ and $\varepsilon\gg t_c$, where transport is determined by charge transfer between QD1 and QD3. The latter region corresponds to $\Gamma_D\gg t_c$, where the coherent system enters the Zeno regime.

The separation between the two regions, defined by $F^{(2)}_{\textrm{I}}/F^{(2)}_{0}=1$ and depicted as a dashed line in Fig.~\ref{fig_mapF}, is determined by the relation
\begin{equation}\label{epsth}
	\varepsilon = \frac{2\sqrt{3}t_c^2}{\sqrt{\Gamma_D^2-8t_c^2}}\,,
\end{equation}
independent of $\Gamma_S$. For fixed coupling $t_c$ and detuning $\varepsilon$, Eq.~(\ref{epsth}) defines a particular value for the coupling to the drain
\begin{equation}\label{gammath}
\Gamma_D^{*} = 2t_c\sqrt{2 + 3(t_c/\varepsilon)^2}
\end{equation}
at which the Fano factor does not allow to discern the nature of charge transport.
To discriminate between coherent and incoherent transport one should therefore choose values of $\Gamma_D$ markedly different from $\Gamma_D^*$. Moreover, Eq.~(\ref{gammath}) determines a crossover between the dominance of coherent and incoherent current fluctuations, tunable by varying $\Gamma_D$ below or above $\Gamma_D^*$.

To have a better understanding of the effects of coherence, we show the Fano factor $F^{(2)}$ and the normalized skewness $F^{(3)}$ for increasing $\Gamma_D$ and different detuning $\varepsilon$ in Figs.~\ref{fig_cumulants}(a) and (b).
In the regime $\Gamma_D \lesssim 2 t_c$, both coherent~(\ref{Fcoh}) and incoherent~(\ref{Fich}) Fano factors are sub-Poissonian and show a local minimum. Coherence suppresses the current fluctuations in agreement with the behavior of a DQD (formed by QD1 and QD3 in this case) having the two energy levels in resonance.\cite{aguado00} The skewness for both the DM in Eq.~(\ref{rhot}) and RE~(\ref{Pt}) exhibit a similar qualitative behavior [cf.~bottom panels in Fig.~(\ref{fig_cumulants})].

In the opposite regime $\Gamma_D\gg t_c$, we obtain interesting and explicit results for the shot noise. Expanding the incoherent Fano factor $F^{(2)}_{\textrm{I}}$ to lowest order in $\Gamma_D^{-1}$ we find
\begin{equation}\label{Fizeno}
F^{(2)}_{\textrm{I}}=\left[1-\frac{2 t_c^2 \varepsilon ^2}{\left(2 t_c^2+\varepsilon ^2\right)^2}\right] \left[1-\frac{8 t_c^4}{\Gamma_S \Gamma_D \left(2 t_c^2+\varepsilon ^2\right)}\right].
\end{equation}
Equation~(\ref{Fizeno}) reveals that $F^{(2)}_I$ is only sub-Poissonian in this limit, as exemplified in Fig.~\ref{fig_cumulants}. In contrast, an expansion of $F^{(2)}_{0}$ in the same coupling limit shows that the coherent Fano factor tends to be super-Poissonian, as we find
\begin{equation}\label{F0zeno}
F^{(2)}_{\textrm{0}}=\left[1+\frac{2 t_c^2 \varepsilon ^2}{\left(2 t_c^2+\varepsilon ^2\right)^2}\right] \left[1-\frac{8 t_c^4}{\Gamma_S \Gamma_D \left(2 t_c^2+\varepsilon ^2\right)}\right].
\end{equation}
Thus, coherence increases fluctuations in the regime of large $\Gamma_D${, in agreement with results for DQDs.\cite{kiesslich06,kiesslich07} Similarly, we find that the coherent third cumulant is larger in absolute value than the incoherent skewness in the limit of large coupling to the drain, as shown in Fig.~\ref{fig_cumulants}.

\begin{figure}[t]
        \includegraphics[width=1\linewidth,clip]{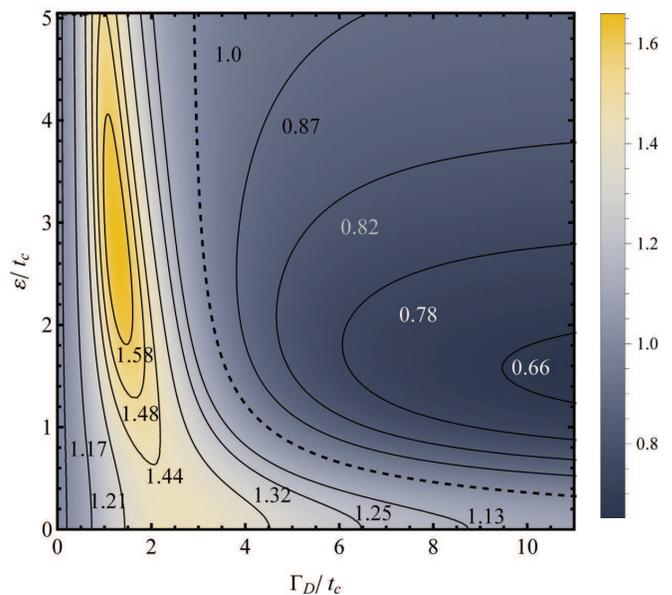}
\caption{Ratio of the incoherent and coherent Fano factors $F^{(2)}_{\mathrm{I}}/F^{(2)}_0$ as a function of the detuning $\varepsilon$ and the coupling to the drain $\Gamma_D$. The separation between enhanced and reduced fluctuations (dashed line) due to coherence is given by Eq.~(\ref{epsth}). ($\Gamma_S/t_c=1/2$).}
\label{fig_mapF}
\end{figure}

\subsection{\label{dephasing}Coherent model and effects of dephasing}

As an intermediate description of the system between the fully coherent and fully incoherent models, we analyze the effects of dephasing on transport through the linear TQD. Exponential damping of the non-diagonal elements of the DM in Eq.~(\ref{rhodt}) yields current fluctuations which differ from the fully coherent DM, Eq.~(\ref{rhot}).

The current in presence of dephasing $I_\phi$ was analyzed in detail in Ref.~\onlinecite{contreras14}; here we briefly present its main characteristics. The average current $I_\phi$ explicitly reads\cite{contreras14}
\begin{equation}\label{currdeph}
	I_\phi = \frac{2et_c^2\Gamma_S\Gamma_D [\gamma\Gamma_\phi^2 + 2t_c^2(4\gamma + \Gamma_D)]}{D_1 + D_2D_3}\,,
\end{equation}
with $\Gamma_\phi = 2\gamma + \Gamma_D$ and furthermore
\begin{eqnarray}\label{currparts}
  D_1 &=& \Gamma_S\Gamma_D\Gamma_\phi(6\gamma + \Gamma_D)\varepsilon^2 \nonumber \,, \\
  D_2 &=& \gamma\Gamma_\phi^2 + 2t_c^2(4\gamma + \Gamma_D) \nonumber\,, \\
  D_3 &=& \Gamma_S\Gamma_D(3\gamma + \Gamma_D) + 2t_c^2(3\Gamma_S + \Gamma_D).
\end{eqnarray}
The current~(\ref{currdeph}) as a function of $\Gamma_D$ and for different dephasing rates $\gamma$ is shown in Figs.~\ref{fig_cumulants}(a) and (b). Compared to the coherent current $I_0$ in Eq.~(\ref{currcoh}), $I_\phi$ is reduced in the regime $\Gamma_D\lesssim 2 t_c$ since dephasing destructs coherent LDT between QD1 and QD3. In the opposite limit $\Gamma_D\gg t_c$, dephasing partially counteracts the coherent trapping of charge between QD1 and QD2 due to the quantum Zeno effect. Consequently, the current is dephasing-enhanced.
The parameter regime for which $I_\phi$ equals the current $I_0$ in Eq.~(\ref{currcoh}) is defined by the relation\cite{contreras14}
\begin{equation}\label{boundary}
	\varepsilon = \sqrt{6}t_c\left[\frac{\gamma(2\gamma + \Gamma_D)^2 + 2t_c^2 (4\gamma + \Gamma_D)}{\Gamma_D(2\gamma + \Gamma_D)^2 - 8t_c^2 (3\gamma + \Gamma_D)}\right]^{1/2}.
\end{equation}
Interestingly, in the limit of small dephasing $\gamma\ll t_c$, Eq.~(\ref{boundary}) reduces to a simpler expression, namely Eq.~(\ref{epsth}) and determines the conditions at which the different models studied here produce the same average current.

As for the current fluctuations, the Fano factor $F^{(2)}_\phi$ and skewness  $F^{(3)}_\phi$ with dephasing are shown in Fig.~\ref{fig_cumulants} for increasing $\Gamma_D$ and for different dephasing rates $\gamma$. In the regime $\Gamma_D\lesssim 2t_c$, it can be seen that the behavior of $F^{(2)}_\phi$ is qualitatively similar to $F^{(2)}_0$ and $F^{(2)}_{\textrm{I}}$. In the opposite limit $\Gamma_D\gg t_c$, we expand $F^{(2)}_\phi$ to lowest order in $\Gamma_D^{-1}$ and find that fluctuations are always sub-Poissonian
\begin{equation}\label{Fdzeno}
F^{(2)}_\phi=1-\left[\Gamma_S \varepsilon ^2+\gamma  \left(\gamma \Gamma_S+4 t_c^2\right)\right]/\gamma \Gamma_S \Gamma_D,
\end{equation}
similarly to the fully incoherent case. Hence, incoherent events involved in transport result in a reduction of the current fluctuations in the regime of large coupling to the drain. A similar behavior is obtained for the skewness since $F^{(3)}_\phi <F^{(3)}_0$ for $\Gamma_D \gg t_c$, as shown in Fig.~\ref{fig_cumulants}.

We finally note that all three models yield essentially the same current and higher-order cumulants for couplings to the drain in the vicinity of $\Gamma_D^*$ defined in Eq.~(\ref{gammath}) [cf.~top to bottom panels in Fig.~\ref{fig_cumulants}]. This result seems to be independent of the physical mechanism that impairs the coherences of the linear TQD. Measurements of high-order cumulants around $\Gamma_D^*$, thus, do not suffice to distinguish between coherent and incoherent descriptions.

\section{Experimental verification}\label{exp}

\begin{figure}[t]
\begin{center}
$\begin{array}{c}
    \multicolumn{1}{l}{\mbox{\bf(a)}}\\
    \epsfig{file=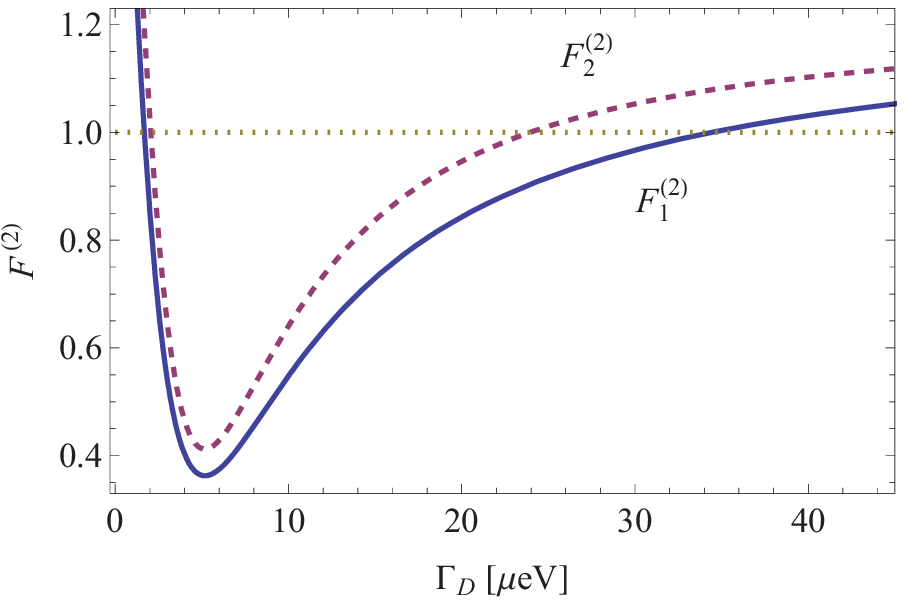,width=0.9\linewidth,clip} \\
       \multicolumn{1}{l}{\mbox{\bf(b)}} \\
    \epsfig{file=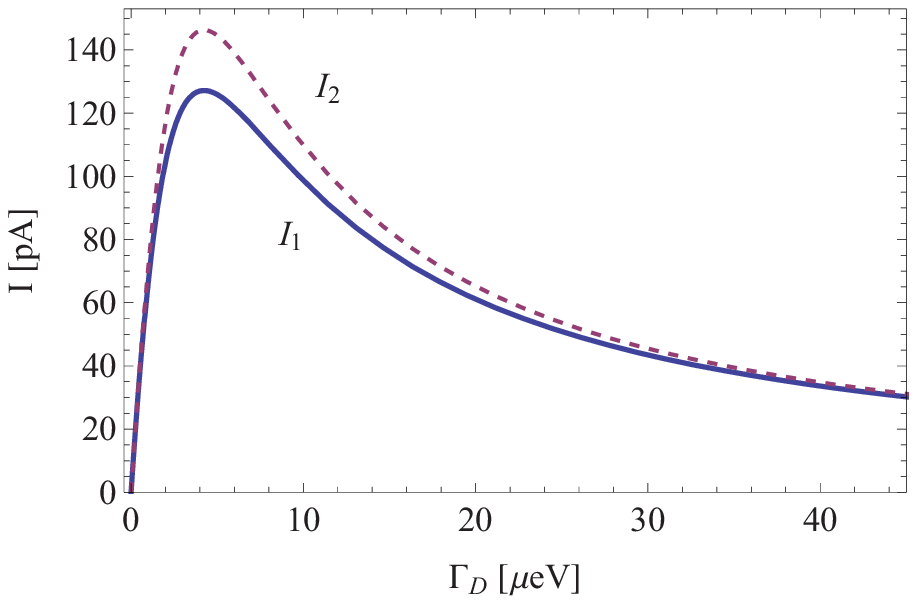,width=0.9\linewidth,clip} \\
\end{array}$
\caption{(a) The Fano factors $F^{(2)}_1$ and $F^{(2)}_2$ for two different values of the coupling to the source $\Gamma_{S,1}=2\mu{\rm eV}$ and $\Gamma_{S,2}=4\mu{\rm eV}$. Measurements of super-Poissonian noise at $\Gamma_D>33\mu{\rm eV}$ and $\Gamma_D>17\mu{\rm eV}$, respectively, reveal that transport is fully coherent. (b) The corresponding currents $I_1$ and $I_2$. (Parameters in the main text).}
\label{fig_exp}
\end{center}
\end{figure}

The sensitivity to coherence exhibited by the high-order cumulants can be used to experimentally verify the presence of quantum coherence in a serially coupled TQD. Coherence is present in the system if measurements of the Fano factor are super-Poissonian in the limit of large coupling to the drain $\Gamma_D\gg t_c$.

In order to determine suitable parameters for detecting super-Poissonian noise, we expand the expression for the coherent Fano factor in Eq.~(\ref{Fcoh}) to lowest order in $\Gamma_D^{-1}$, Eq.~(\ref{F0zeno}), and then we find the conditions for which $F^{(2)}_0>1$, yielding the relation
\begin{equation}\label{Fexp}
\Gamma_S \Gamma_D > \frac{4 t_c^2\left(4 t_c^4+6 t_c^2\varepsilon^2+\varepsilon^4\right)}{\varepsilon^2\left(2 t_c^2+\varepsilon^2\right)}.
\end{equation}
This is in agreement with the enhanced noise owing to coherence observed in a DQD with finite detuning and asymmetric couplings to the reservoirs.\cite{haug06,kiesslich07}

Considering representative parameters for lateral QDs,\cite{hayashi03,busl13,vandersypen13} we conceive the following experimental scheme: The current and Fano factor are measured with two different values of the incoherent coupling to the source, $\Gamma_{S,1}=2\mu{\rm eV}$ and $\Gamma_{S,2}=4\mu{\rm eV}$. In both realizations the energy detuning $\varepsilon=6\mu{\rm eV}$ and the interdot coupling $t_c=3\mu{\rm eV}$ are kept constant, while the coupling of QD3 to the drain $\Gamma_D$ is varied in the range $0-50\mu{\rm eV}$. We select these parameters to ensure that the Zeno regime is reached and that the detected Fano factors are super-Poissonian.
In accord with expression~(\ref{Fexp}), the measured fluctuations $F^{(2)}_{1}$ and $F^{(2)}_{2}$ are expected to be super-Poissonian for $\Gamma_D>33\mu{\rm eV}$ and $\Gamma_D>17\mu{\rm eV}$, respectively, as shown in Fig.~\ref{fig_exp}(a). Observations of sub-Poissonian noise in this regime reflects therefore the occurrence of a physical mechanism which brings the TQD into a classical, incoherent regime. We verify that the coupling to the drain for incoherent transport $\Gamma_D^*$, Eq.~(\ref{gammath}), is far from the relevant values of $\Gamma_D$ to detect super-Poissonian noise.  For our parameters, this value corresponds to $\Gamma_D^*\cong 10\mu{\rm eV}$.

For large coupling to the drain, the current~(\ref{currcoh}) obtained with the coherent DM tends to be suppressed, cf.~Fig.~\ref{fig_exp}(b). Nevertheless, the chosen parameters predict values of the measured currents $I_{1}$ and $I_{2}$ detectable in present technologies for arrays of lateral QDs. In addition, increased currents with respect to Eq.~(\ref{currcoh}) can be used to evidence the effect of decaying coherences since the model including pure dephasing predicts enhanced transport for sufficiently large $\Gamma_D$ and finite detuning $\varepsilon$.\cite{contreras14}

\section{\label{conc}Conclusions}

We have analyzed coherent and incoherent FCS of a serially coupled TQD in a transport configuration, and addressed the question on how to discern whether transport is due to quantum coherence or to incoherent processes.

A density matrix approach was used to model the coherent tunneling of charge between electronic sites, while incoherent transport was defined in terms of a rate equation. The effects of pure dephasing were also included in order to consider an intermediate description with decaying coherences of the density matrix.

Our analytical results reveal that the full coherent model and the rate equation yield the same current, while higher-order cumulants (shot noise and skewness) differ, demonstrating that transport in the TQD is sensitive to coherence. In particular, coherence enhances the current fluctuations for sufficiently large coupling to the drain and finite energy detuning, resulting in super-Poissonian noise. In contrast, the rate equation and the description with dephasing yield sub-Poissonian Fano factors. Therefore measures of the Fano factor in this coupling limit can be used to evidence the occurrence of coherence in charge transport across the TQD.

We furthermore found the conditions for which it is not possible to discern the nature of transport since the three models yield basically the same statistics. This regime can be reached by varying a single external parameter, namely the incoherent coupling to the drain. Our findings are relevant in order to examine the effects of coherence in larger arrays of QDs (e.g.~the recently demonstrated quadruple quantum dot\cite{tarucha14b}) or generally in other transport systems with controllable parameters.

\section{\label{ack}Acknowledgements}

The authors are grateful to S.~F. Huelga and M.~B. Plenio for enlightening discussions and for their critical reading of the manuscript. Fruitful discussions with R. Aguado and G. Platero are also acknowledged. This work was supported by DGAPA-UNAM (Mexico) through project PAPIIT IA101416, the ERC Synergy grant BioQ, the EU projects SIQS and DIADEMS and the Alexander von Humboldt Foundation.

\appendix

\section{\label{appss}Derivation of the rate equation for the linear TQD}

We derive a Pauli rate equation for the occupation probabilities of the electronic states of the linear TQD $\mathbf{P}=(\rho_{11},\rho_{22},\rho_{33},\rho_{44})^{T}$, starting from the time evolution of the fully coherent DM, Eq.~(\ref{rhot}). Explicitly, the equations of motion of the populations of the DM read
\begin{eqnarray}\label{pop0}
\dot{\rho_{11}}(t)&=& i  t_c [\rho_{12}(t)-\rho_{21}(t)]+\Gamma_S \rho_{00}(t) \nonumber\,, \\
\dot{\rho_{22}}(t)&=& -i  t_c [\rho_{12}(t)-\rho_{21}(t)-\rho_{23}(t)+\rho_{32}(t)] \nonumber\,,\\
\dot{\rho_{33}}(t)&=& -i  t_c [\rho_{23}(t)-\rho_{32}(t)]-\Gamma_D \rho_{33}(t) \nonumber\,,\\
\dot{\rho_{00}}(t)&=& -\Gamma_S \rho_{00}(t)+\Gamma_D \rho_{33}(t),
\end{eqnarray}
and for the coherences we have
\begin{eqnarray}\label{coh0}
\dot{\rho_{12}}(t)&=& i t_c \rho_{13}(t) +i t_c[\rho_{11}(t)-\rho_{22}(t)] -i \varepsilon \rho_{12}(t) \nonumber\,,\\
\dot{\rho_{23}}(t)&=& -i t_c \rho_{13}(t) +i t_c [\rho_{22}(t)-\rho_{33}(t)] \nonumber \\
&&+\left(i\varepsilon-\Gamma_D/2\right)\rho_{23}(t) \nonumber \,, \\
\dot{\rho_{13}}(t)&=& i t_c [\rho_{12}(t)-\rho_{23}(t)]-\Gamma_D \rho_{13}(t)/2.
\end{eqnarray}

As we are interested in the steady-state properties of the TQD, the rate equation (RE) is calculated by setting to zero the time derivatives of the coherences from Eqs.~(\ref{coh0}), $\dot{\rho}_{ij}^s=0$ (where the superscript ``s'' refers to the steady-sate). Then, the set of algebraic equations for $\rho_{ij}^s$ is solved and the results are substituted in the equations of motion for the occupation probabilities $\dot{\rho}_{ii}(t)$, (\ref{pop0}). As a result, we obtained the RE~(\ref{Pt}), with the generator $\mathcal{L}_{\mathrm{I}}$ given in Eq.(\ref{Lich}) of the main text.

This rate equation approximation is essentially exact for the steady-state since the population differences occurring in the time evolution of the coherences (\ref{coh0}) are constant.\cite{sargent,milburn} Moreover, the regime for large incoherent rate to the drain $\Gamma_D\gg t_c$ with finite $\varepsilon$ is the most relevant in our work, making the coherences to decay faster than the populations difference. Thus, the Pauli RE~(\ref{Pt}) yields reliable results, provided that we restrict ourselves to the steady-state transport in the TQD.

\section{\label{appFCS}Counting statistics for the linear TQD}

To determine the zero-frequency counting statistics of the serially coupled TQD, we first introduce the counting variable $\xi$ by the substitution $d_3\rho d^{\dagger}_3\rightarrow e^{\xi}d_3\rho d^{\dagger}_3$ in the models studied here, equations~(\ref{rhot}),~(\ref{Pt}) and~(\ref{rhodt}). We thereby transform the generators of the dynamics into the deformed generators $\mathcal{L}_i\rightarrow\mathcal{L}_{i,\xi}$, with $i=0,I,\phi$. Then we apply the characteristic polynomial approach to analytically calculate the cumulants of the distribution $p(N)$. We refer to Ref.~\onlinecite{bruderer14} for the details related to this method.

\subsection{\label{appdeph}Density Matrix approach: Full coherent model and pure dephasing}

The FCS for the full coherent model~(\ref{rhot}) and for the description including dephasing~(\ref{rhodt}) is obtained from the deformed generators $\mathcal{L}_{0,\xi}$ and $\mathcal{L}_{\phi,\xi}$, respectively. Both of them are expressed as matrices of dimension $16\times16$ in the Liouville space.

The analytical result for the current obtained with the coherent DM~(\ref{rhot}) is given explicitly in the main text, cf.~Eq.~(\ref{currcoh}). The second cumulant reads
\begin{equation}\label{c2phi}
	c_{2} = \frac{A_1}{A_0^3},
\end{equation}
with
\begin{eqnarray}\label{c2parts1}
  A_0&=&4 t_c^4 \left(\Gamma _D+3 \Gamma _S\right)+2 t_c^2 \Gamma _D^2 \Gamma _S+\varepsilon ^2 \Gamma _D^2 \Gamma _S \nonumber \,,\\
  A_1&=& 4 t_c^4 \Gamma _D \Gamma _S[16 t_c^8 \Gamma _S \left(\Gamma _D^2+11 \Gamma _S^2\right)\nonumber \\
  &&+\varepsilon ^2 \Gamma _D^4 \Gamma _S^3 \left(6 t_c^2+\varepsilon ^2\right)\nonumber \\
 &&+4 t_c^4 \Gamma _S^3 \left(\Gamma _D^2-4 \varepsilon ^2\right) \left(\Gamma _D^2-2 t_c^2\right)].
\end{eqnarray}	
The Fano factor for the coherent model is calculated directly from the cumulants as $F^{(2)}=c_2/c_1$, cf.~Eq.~(\ref{Fcoh}) in the main text.

For the coherent third cumulant we find
\begin{equation}\label{c3}
	c_{3} = \frac{4 t_c^4 \Gamma _D \Gamma _S}{A_{0}^{5}}\sum_{i=1}^{8}A_i,
\end{equation}
where the $A_i$ are
\begin{eqnarray}\label{c3longparts1}
  A_{1} &=& 256 t_c^{16} [\Gamma _D^4-6 \Gamma _D^3 \Gamma _S+66 \Gamma _D^2 \Gamma _S^2\nonumber \\
  &&-90 \Gamma _D \Gamma _S^3+93 \Gamma_S^4]\nonumber \,, \\
  A_{2}&=&128 t_c^{14} \Gamma _S [-12 \varepsilon ^2 \Gamma _S \left(2 \Gamma _D+\Gamma _S\right) \left(5 \Gamma _S-\Gamma _D\right) \nonumber \\
  && -\Gamma _D^2 \left(2 \Gamma _D^3+6 \Gamma _D^2 \Gamma _S+21 \Gamma _D \Gamma _S^2+219 \Gamma _S^3\right)] \nonumber\,,\\
   A_{3}&=& 128 t_c^{12} \Gamma _S [3 \Gamma _D^4 \Gamma _S \left(\Gamma _D^2+8 \Gamma _S^2\right) \nonumber \\
  &&-\varepsilon ^2 \Gamma _D^2 \left(\Gamma _D^3+12 \Gamma _D^2 \Gamma _S-3 \Gamma _D \Gamma _S^2+234 \Gamma _S^3\right)\nonumber\\
  &&-12 \varepsilon ^4 \Gamma _S^2 \left(\Gamma _D+\Gamma _S\right)]\nonumber \,, \\
  A_{4}&=&32 t_c^{10} \Gamma _D^2 \Gamma _S^2 [12 \varepsilon ^4 \Gamma _S \left(\Gamma _D-16 \Gamma _S\right)\nonumber \\
  &&-2 \Gamma _D^4 \Gamma _S \left(\Gamma _D+6 \Gamma _S\right)\nonumber \\
  &&+3 \varepsilon ^2 \Gamma _D^2 \left(6 \Gamma _D^2+10 \Gamma _D \Gamma _S+107 \Gamma _S^2\right)] \nonumber\,, \\
  A_{5}&=&16 t_c^8 \Gamma _D^2 \Gamma _S^2 [6 \varepsilon ^4 \Gamma _D^2 \left(\Gamma _D^2+\Gamma _D \Gamma _S+37 \Gamma _S^2\right)\nonumber \\
  &&+\Gamma _D^6 \Gamma _S^2-24 \varepsilon ^6 \Gamma _S^2-12 \varepsilon ^2 \Gamma _D^4 \Gamma _S \left(\Gamma _D+4 \Gamma _S\right)] \nonumber \,, \\
A_{6}&=&8 \varepsilon ^2 t_c^6 \Gamma _D^4 \Gamma _S^3 [\Gamma _S \left(4 \Gamma _D^4-69 \varepsilon ^2 \Gamma _D^2+36 \varepsilon ^4\right)\nonumber \\
&&-9 \varepsilon ^2 \Gamma _D^3]\nonumber \,, \\
  A_{7}&=&8 \varepsilon ^4 t_c^4 \Gamma _D^6 \Gamma _S^3 \left[6 \Gamma _D^2 \Gamma _S-\varepsilon ^2 \left(\Gamma _D+12 \Gamma _S\right)\right]\nonumber \,, \\
  A_{8}&=&\varepsilon ^6 \Gamma _D^8 \Gamma _S^4 \left(14 t_c^2+\varepsilon ^2\right).
\end{eqnarray}

We turn now to the description of the TQD including pure-dephasing. The current is given explicitly in the main text, see Eq.~(\ref{currdeph}). The second cumulant, related to the shot noise, was found to be\cite{contreras14}
\begin{equation}\label{c2phi}
	c_{2,\phi} = \frac{B_1}{B_9}\bigg(B_2B_3 + \sum_{i=4}^8 B_i\bigg),
\end{equation}
with the $B_i$ given by the expressions
\begin{eqnarray}\label{c2longparts1}
  B_1 &=& 2\Gamma_S \Gamma_D t_c^2 [\gamma\Gamma_\phi^2 + 2 t_c^2 (4 \gamma +\Gamma_D)]\nonumber\,, \\
  B_2 &=& 2\Gamma_S^2 \Gamma_D t_c^2\Gamma_\phi^2\nonumber\,, \\
  B_3 &=& \gamma\Gamma_\phi(40 \gamma^4+92\gamma^3 \Gamma_D +60 \gamma^2 \Gamma_D^2 +17 \gamma \Gamma_D^3+2 \Gamma_D^4) \nonumber \\
	&&+\varepsilon^2 (80 \gamma^4+264 \gamma^3 \Gamma_D+164 \gamma^2 \Gamma_D^2+34 \gamma  \Gamma_D^3+3 \Gamma_D^4) \nonumber \\
    &&+8 \gamma \Gamma_D \varepsilon^4 \nonumber\,, \\
  B_4 &=& 16t_c^8 (4 \gamma +\Gamma_D)^2 [2 \gamma(9 \Gamma_S^2+\Gamma_D^2)+11 \Gamma_S^2 \Gamma_D+\Gamma_D^3]\nonumber\,,
\end{eqnarray}
and furthermore
\begin{eqnarray}\label{c2longparts2}
  B_5 &=& \Gamma_S^2 \Gamma_D^2\Gamma_\phi^3[\varepsilon^4 (28 \gamma^2+8\gamma \Gamma_D+\Gamma_D^2)+\gamma^2\Gamma_\phi^2 (\Gamma_D^2)\nonumber \\
  &&+\gamma^2\Gamma_\phi^2 (7 \gamma^2+5 \gamma  \Gamma_D)\nonumber \\
  &&+\gamma\varepsilon^2\Gamma_\phi(28 \gamma^2+14\gamma\Gamma_D+\Gamma_D^2) ] \nonumber\,, \\
  B_6 &=& 8t_c^6 \Gamma_\phi(4 \gamma +\Gamma_D)[8 \gamma^3(9 \Gamma_S^2+\Gamma_D^2 )\nonumber \\
  &&+8 \gamma^2  (15 \Gamma_S^2 \Gamma_D+\Gamma_D^3) \nonumber \\
	&&+2\gamma(14 \Gamma_S^2 \Gamma_D^2+\Gamma_D^4 )-\Gamma_S^2 \Gamma_D  (\Gamma_D^2-4 \varepsilon^2 )] \nonumber\,, \\
  B_7 &=& 4t_c^4\Gamma_\phi[16\gamma^6  (9 \Gamma_S^2+\Gamma_D^2 ) + 16\gamma^5(39 \Gamma_S^2 \Gamma_D+2 \Gamma_D^3) \nonumber\\
	&&+24 \gamma^4  (30 \Gamma_S^2 \Gamma_D^2+\Gamma_D^4 )+\Gamma_S^2 \Gamma_D^4  (\Gamma_D^2-4\varepsilon^2)] \nonumber\,, \\
 B_8 &=& 4t_c^4\Gamma_\phi\{4\gamma^3[\Gamma_S^2(85\Gamma_D^3+44 \Gamma_D \varepsilon^2 )+2\Gamma_D^5] \nonumber\\
	&&+\gamma^2[\Gamma_S^2  (78 \Gamma_D^4+88 \Gamma_D^2 \varepsilon ^2 )+\Gamma_D^6]\nonumber \\
&&+\gamma\Gamma_S^2 \Gamma_D^3  (11 \Gamma_D^2+4 \varepsilon^2)\} \nonumber\,, \\
  B_9 &=& \Gamma_\phi\{\Gamma_S \Gamma_D\Gamma_\phi\varepsilon^2 (6 \gamma +\Gamma_D)\nonumber \\
    &&+ [\gamma\Gamma_\phi^2+2 t_c^2 (4 \gamma +\Gamma_D)][\Gamma_S \Gamma_D (3 \gamma +\Gamma_D) \nonumber\\
	&&+2 t_c^2 (3 \Gamma_S+\Gamma_D)]\}^3,
\end{eqnarray}
where $\Gamma_\phi = 2\gamma + \Gamma_D$.

The Fano factor with dephasing $F^{(2)}_{\phi}=c_{2,\phi}/c_{1,\phi}$, takes the explicit form
\begin{equation}\label{Fphi}
	F^{(2)}_{\phi} =1-\frac{\Gamma_S \Gamma_D}{\Gamma _{\phi } (D_1+D_2 D_3)^2} \sum_{j=10}^{16} B_j.
\end{equation}
The $D_i$ are defined in Eq.~(\ref{currparts}) of the main text, and for the $B_j$ we have
\begin{eqnarray}\label{Fdlongparts1}
  B_{10} &=& 2 \varepsilon ^2 t_c^2 \Gamma _{\phi }^2 [2 \gamma  \Gamma _{\phi }^2 \Gamma _D \left(6 \gamma +\Gamma _D\right)+\Gamma _S (64 \gamma ^4+48 \gamma ^3 \Gamma _D\nonumber\\
  &&+4 \gamma ^2 \Gamma _D^2-2 \gamma  \Gamma _D^3-\Gamma _D^4)]\nonumber\,, \\
  B_{11} &=& 2 \gamma  t_c^2 \Gamma _{\phi }^2 [\gamma  \Gamma _{\phi }^2 (2 \Gamma _D^3+\Gamma _D^2 \left(10 \gamma +9 \Gamma _S\right)\nonumber\\
  &&+6 \gamma  \Gamma _D \left(2 \gamma +5 \Gamma _S\right)+16 \gamma ^2 \Gamma _S)-8 \varepsilon ^4 \Gamma _D \Gamma _S]\nonumber\,, \\
	B_{12} &=& \gamma  \left(\gamma ^2+\varepsilon ^2\right) \Gamma _{\phi }^4 \Gamma _D \Gamma _S \left(\Gamma _{\phi }^2+4 \varepsilon ^2\right) \nonumber\,, \\
	B_{13} &=& 8 \varepsilon ^2 t_c^4 \Gamma _{\phi } [8 \gamma ^2 \Gamma _D \left(6 \gamma +11 \Gamma _S\right)+\Gamma _D^3 \left(12 \gamma +5 \Gamma _S\right) \nonumber \\
    &&+2 \gamma  \Gamma _D^2 \left(22 \gamma +17 \Gamma _S\right)+56 \gamma ^3 \Gamma _S+\Gamma _D^4] \nonumber\,, \\
	B_{14} &=& 4 \gamma  t_c^4 \Gamma _{\phi }^2 [24 \gamma ^2 \Gamma _D \left(7 \gamma +9 \Gamma _S\right)+3 \Gamma _D^3 \left(14 \gamma +5 \Gamma _S\right) \nonumber \\
    &&+2 \gamma  \Gamma _D^2 \left(70 \gamma +53 \Gamma _S\right)+24 \gamma ^3 \left(2 \gamma +5 \Gamma _S\right)+4 \Gamma _D^4]  \nonumber\,, \\
	B_{15} &=& 32 t_c^8 \left(4 \gamma +\Gamma _D\right){}^2 \left(6 \gamma +3 \Gamma _D-\Gamma _S\right) \nonumber\,, \\
	B_{16} &=& 8 t_c^6 \Gamma _{\phi } \left(4 \gamma +\Gamma _D\right) [48 \gamma ^3+2 \Gamma _D \left(4 \gamma +\Gamma _D\right) \left(9 \gamma +\Gamma _D\right) \nonumber \\
    &&+\Gamma _S \left(24 \gamma ^2+32 \gamma  \Gamma _D+7 \Gamma _D^2-4 \varepsilon ^2\right)].
\end{eqnarray}

The analytical expression for the third cumulant including the effects of dephasing $c_{3,\phi}$, is extremely lengthy and therefore it is not shown explicitly. Its physical content is explored in subsection~\ref{dephasing} and Fig.~(\ref{fig_cumulants}) in the main text.

\subsection{\label{apptransport}Fully incoherent model}

For incoherent transport we use the deformed generator $\mathcal{L}_{I,\xi}$. Since it corresponds to a description without coherences, $\mathcal{L}_{I,\xi}$ has dimensions of $4\times4$. As explained in the main text, the first cumulant obtained with both, coherent and incoherent descriptions equal each other, cf.~Eq.~(\ref{currcoh}). The incoherent second cumulant reads
\begin{equation}\label{c2ic}
c_{2,I}=\frac{4 t_c^4 \Gamma _D \Gamma _S}{A_0^3}E_{1},
\end{equation}
with $E_{1}=16 t_c^8 \left(\Gamma _D^2+11 \Gamma _S^2\right)+4 t_c^4 \Gamma _D^2 \Gamma _S^2 \left(10 t_c^2+\Gamma _D^2\right)$ and $A_0$ is defined in Eq.~(\ref{c2parts1}).

For the incoherent skewness we have
\begin{equation}\label{c3ic}
c_{3,I}=\frac{4 t_c^4 \Gamma _D \Gamma _S}{A_0^5}\sum_{i=2}^{8}E_i\,,
\end{equation}
with the functions $E_i$ given by
\begin{eqnarray}\label{c3iclongparts1}
E_{2}&=&256 t_c^{16} [\Gamma _D^4-6 \Gamma _D^3 \Gamma _S+66 \Gamma _D^2 \Gamma _S^2\nonumber \\
&&-78 \Gamma _D \Gamma _S^3+129 \Gamma _S^4]\nonumber \,, \\
E_{3}&=&128 t_c^{14} \Gamma _S [12 \varepsilon ^2 \Gamma _S \left(2 \Gamma _D^2-5 \Gamma _D \Gamma _S+7 \Gamma _S^2\right) \nonumber \\
&&+\Gamma _D^2 \left(3 \Gamma _S \left(10 \Gamma _D^2+33 \Gamma _S^2\right)-2 \Gamma _D^3-51 \Gamma _D \Gamma _S^2\right)] \nonumber \,,\\
E_{4}&=&64 t_c^{12} \Gamma _S [-2 \varepsilon ^2 \Gamma _D^5+6 \Gamma _D^4 \Gamma _S \left(\Gamma _D^2+4 \varepsilon ^2\right)\,, \\
E_{5}&=&32 t_c^{10} \Gamma _D^2 \Gamma _S^2 [3 \Gamma _S^2 \left(17 \varepsilon ^2 \Gamma _D^2+3 \Gamma _D^4-8 \varepsilon ^4\right)\nonumber \\
&&+6 \varepsilon ^2 \Gamma _D^4-2 \Gamma _D \Gamma _S \left(9 \varepsilon ^2 \Gamma _D^2+\Gamma _D^4+6 \varepsilon ^4\right)] \nonumber \,, \\
E_{6}&=&16 t_c^8 \Gamma _D^2 \Gamma _S^2 [6 \varepsilon ^4 \Gamma _D^2 \left(\Gamma _D^2-\Gamma _D \Gamma _S+9 \Gamma _S^2\right)\nonumber \\
&&-3 \varepsilon ^2 \Gamma _D^4 \Gamma _S \left(\Gamma _D-4 \Gamma _S\right)+\Gamma _D^6 \Gamma _S^2-24 \varepsilon ^6 \Gamma _S^2] \nonumber \,, \\
E_{7}&=&4 \varepsilon ^2 t_c^4 \Gamma _D^4 \Gamma _S^3 [3 \varepsilon ^2 \Gamma _D^4 \Gamma _S-2 \varepsilon ^4 \Gamma _D^2 \left(\Gamma _D-6 \Gamma _S\right)\nonumber \\
&&+2 t_c^2 \left(-3 \varepsilon ^2 \Gamma _D^2 \left(\Gamma _D-7 \Gamma _S\right)+\Gamma _D^4 \Gamma _S+12 \varepsilon ^4 \Gamma _S\right)] \nonumber \,, \\
E_{8}&=&\varepsilon ^6 \Gamma _D^8 \Gamma _S^4 \left(2 t_c^2+\varepsilon ^2\right).
\end{eqnarray}

\section{\label{appcoh}Elements of the steady-state DM}

\begin{figure}[t]
\begin{center}
$\begin{array}{c}
\multicolumn{1}{l}{\mbox{\bf(a)}}\\
    \epsfig{file=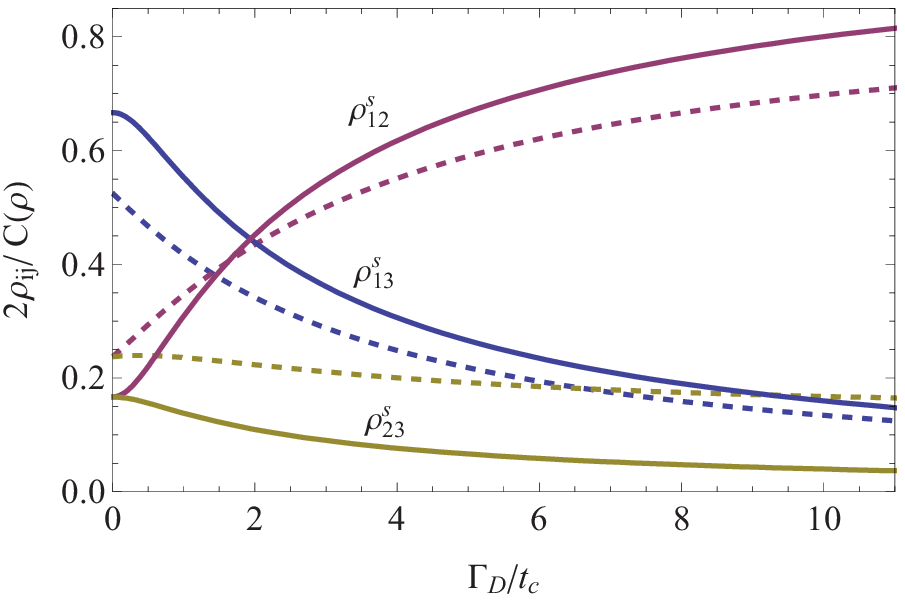,width=0.9\linewidth,clip} \\
       \multicolumn{1}{l}{\mbox{\bf (b)}} \\
        \epsfig{file=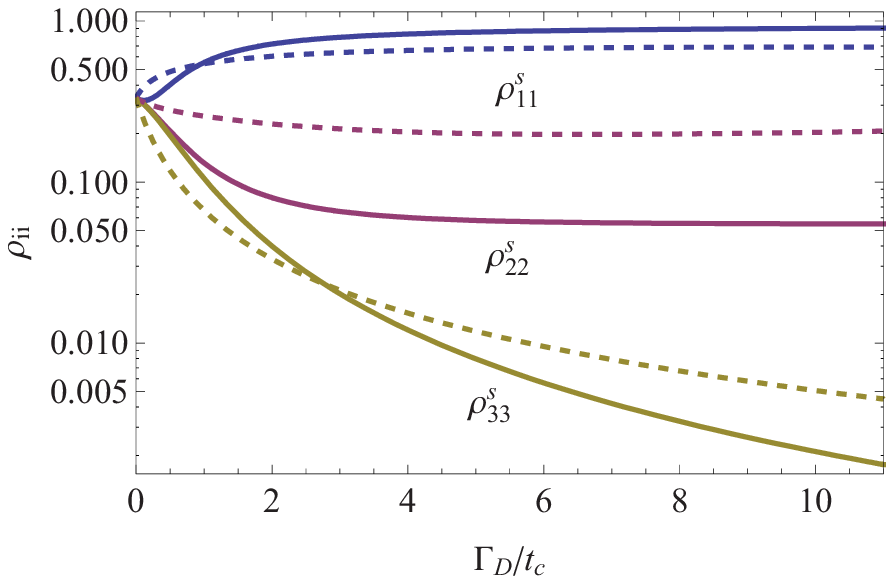,width=0.9\linewidth,clip} \\
\end{array}$
\caption{(a) Coherences $|\rho^s_{ij}|$ weighted by the overall coherence $C(\rho)$, and (b) occupation probabilities of the steady-state DM for the fully coherent model (solid line) and for the description including pure dephasing (dashed). Parameters: $\varepsilon/t_c=4$, $\gamma/t_c=2/3$ and $\Gamma_S/t_c=1/2$.}
\label{fig_rhoij}
\end{center}
\end{figure}

The steady-state of the linear TQD described by the DM approach, $\rho^{\textrm{s}}$ is determined by solving $\dot{\rho}=\mathcal{L}_i\rho^{\textrm{s}}=0$, with $i=0,\phi$. We briefly discuss here the behavior of the steady-state elements of the full coherent DM~(\ref{rhot}) and the model including pure dephasing,~(\ref{rhodt}). The former is obtained by setting to zero the dephasing rate, $\gamma=0$, in the expressions below.

The occupation probabilities read\cite{contreras14}
\begin{eqnarray}\label{eqs_pop}
\rho_{11}^s &=& \frac{\Gamma_S}{2}\,\frac{2t_c^2 F_{1} + (4\gamma+\Gamma _D)[8t_c^4 +\Gamma_{\phi}\Gamma_D(\gamma\Gamma_{\phi}+2\varepsilon^2)]}{D_1+D_2D_3} \nonumber \,,\\
\rho_{22}^s &=&\frac{\Gamma_S}{2}\,\frac{2t_c^2\Gamma_{\phi} F_{2}+ 8 t_c^4(4\gamma +\Gamma_D)+\gamma\Gamma_{\phi}\Gamma_D(\Gamma_{\phi}^2+4\varepsilon ^2)}{D_1+D_2D_3}\nonumber \,,\\
\rho_{33}^s &=&\frac{\Gamma_S}{2}\,\frac{4t_c^2[\gamma\Gamma_{\phi}^2+2t_c^2(4\gamma +\Gamma_D)]}{D_1+D_2D_3} \,,
\end{eqnarray}
with $\Gamma_\phi = 2\gamma + \Gamma_D$ and the $D_i$ are defined in Eq.~(\ref{currparts}). We also used the abbreviations $F_{1}=8\gamma^3 + 24\gamma^2\Gamma_D + 10\gamma\Gamma_D^2 + \Gamma_D^3$ and $F_{2}=4\gamma^2+6\gamma\Gamma_D+\Gamma_D^2$.
The occupation of the empty state is obtained as $\rho_{00}^s=1-\sum_{i=1}^{3}\rho_{ii}^{s}$. Note that in the Coulomb blockade regime and infinite bias voltage, the steady-state current is given by $I=e\Gamma_D\rho_{33}^s$.

The coherences were found to be\cite{contreras14}
\begin{eqnarray}\label{eqs_coh}
\rho_{12}^{\rm s} &=& \frac{t_c \Gamma _D\Gamma _S [(\varepsilon + i\gamma)\Gamma_{\phi}^2 + 2i t_c^2(4 \gamma + \Gamma _D)]}{D_1 + D_2 D_3} \nonumber \,, \\
\rho_{23}^{\rm s} &=& \frac{i t_c\Gamma_D\Gamma_S [2t_c^2(4 \gamma +\Gamma_D) + \gamma\Gamma_{\phi}(\Gamma_{\phi} + 2i\varepsilon)]}{D_1 + D_2 D_3} \nonumber \,,\\
\rho_{13}^{\rm s} &=& \frac{2i\varepsilon t_c^2 \Gamma_D \Gamma_S (4\gamma + \Gamma_D)}{D_1 + D_2 D_3}.
\end{eqnarray}
The weights of the non-diagonal elements $\rho_{ij}^{s}$ with respect to the overall coherence in the system, defined as~\cite{baumgratz14} $C(\rho) = \sum_{i\neq j} |\rho_{ij}|$, are quantified in the form $2|\rho_{ij}^{s}|/C(\rho^{s})$. This ratio and the occupations~(\ref{eqs_pop}) with and without dephasing are shown in Fig.~\ref{fig_rhoij}(a) and~(b), respectively, as a function of $\Gamma_D$.

For the fully coherent model ($\gamma=0$) we observe that $\rho^s_{13}$, revealing LDT between QD1 and QD3, accounts for most of the coherence in the regime $\Gamma_D\lesssim 2t_c$. At this coupling, the occupation $\rho_{33}^s$ of QD3 is large and decreases for increasing $\Gamma_D$, in accord to the behavior of the current, see Fig.~(\ref{fig_cumulants}). The coherence $\rho^s_{12}$ is the dominating contribution to the overall coherence for $\Gamma_D\gg t_c$, as a signature of the Zeno effect; in this regime the elements $\rho_{11}^s$ and $\rho_{22}^s$ are approximately constant and account for most of the occupation of the TQD, cf.~Fig.~\ref{fig_rhoij}(b).

Pure dephasing generally impairs the coherence of the TQD. Hence, the strength of $\rho^s_{13}$ is reduced due to dephasing for small $\Gamma_D$, Fig.~\ref{fig_rhoij}(a), with the corresponding reduction of the current and of $\rho_{33}^s$. For sufficiently large $\Gamma_D$, dephasing partially alleviates the charge localization resulting in a reduction of $\rho^s_{12}$. Consequently, the occupation $\rho_{33}^s$ is dephasing-enhanced for $\Gamma_D\gg t_c$ while the difference between $\rho_{22}^s$ and $\rho_{11}^s$ is reduced, cf.~Fig.~\ref{fig_rhoij}(b).


\section{\label{appich}Incoherent rates between electronic states $\Gamma_{ij}$}
\begin{figure}[b]
        \includegraphics[width=0.95\linewidth,clip]{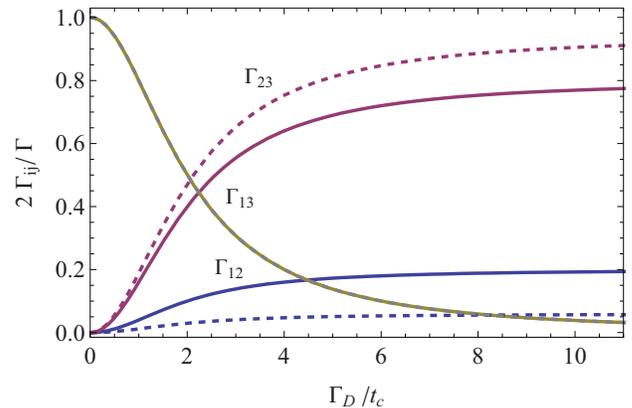}
    \caption{Incoherent transition rates between electronic states of the TQD $\Gamma_{ij}$, for $\varepsilon/t_c=2$ (solid line) and $\varepsilon/t_c=4$ (dashed line). We used $\Gamma_S/t_c=1/2$. }
    \label{fig_rates}
\end{figure}

We turn to analyze the incoherent transition rates between electronic states of the TQD, $\Gamma_{ij}$ defined in Eqs.~(\ref{rates_ich}) in the main text. Similarly to the DM approach, we assess the effect of the rates by defining the sum $\Gamma=\sum_{i\neq j}\Gamma_{ij}$. The weighted contribution of the specific rates, accounted as $2 \Gamma_{ij}/\Gamma$, are presented in Fig.~\ref{fig_rates} as a function of $\Gamma_D$ and for different values of the energy detuning $\varepsilon$. We observe that the rate $\Gamma_{13}$ is the dominating contribution to the total transitions in the regime $\Gamma_D\lesssim 2 t_c$, and decays for increasing $\Gamma_D$. In addition, its weight is not strongly sensitive to the detuning of the central QD, unlike the LDT exhibited by the coherent description of the TQD.
The incoherent rates between adjacent sites, $\Gamma_{12}$ and $\Gamma_{23}$, account for most of the transitions for sufficiently large $\Gamma_D$, indicating sequential transport of charge in this limit.


\end{document}